# Accurate and efficient predictions of keyhole dynamics in laser materials processing using machine learning-aided simulations

Jiahui Zhang [a], Runbo Jiang [b], Kangming Li [a], Pengyu Chen [a], Shengbo Bi [a], Xiao Shang [a], Zhiying Liu[a], Jason Hattrick-Simpers [a], Brian J. Simonds [c], Qianglong Wei [d], Hongze Wang [d], Tao Sun [e], Anthony D. Rollett [f], Yu Zou [a,*]

[a] *Department of Materials Science and Engineering, University of Toronto, Toronto, ON M5S 3E4, Canada*

[b] *Advanced Light Source (ALS) Division, Lawrence Berkeley National Laboratory, Berkeley, CA 94720, USA*

[c] *Applied Physics Division, National Institute of Standards and Technology, Boulder, CO 80305, USA*

[d] *School of Materials Science & Engineering, Shanghai Jiao Tong University, Shanghai, 200240, China*

[e] *Department of Mechanical Engineering, Northwestern University, Evanston, IL 60208, USA*

[f] *Department of Materials Science and Engineering, Carnegie Mellon University, Pittsburgh, PA 15213, USA*

* Corresponding author. Email: mse.zou@utoronto.ca (Y. Z.)

## Abstract

The keyhole phenomenon has been widely observed in laser materials processing, including laser welding, remelting, cladding, drilling, and additive manufacturing. Keyhole-induced defects, primarily pores, dramatically affect the performance of final products, impeding the broad use of these laser-based technologies. The formation of these pores is typically associated with the dynamic behavior of the keyhole. So far, the accurate characterization and prediction of keyhole features, particularly keyhole depth, as a function of time, has been a challenging task. In situ characterization of keyhole dynamic behavior using the synchrotron X-ray technique is informative but complicated and expensive. Current simulations are generally hindered by their poor accuracy and generalization abilities in predicting keyhole depths due to the lack of accurate laser absorptance data. In this study, we develop a machine learning-aided simulation method that accurately predicts keyhole dynamics, especially in keyhole depth fluctuations, over a wide range of processing parameters. In two case studies involving titanium and aluminum alloys, we achieve keyhole depth prediction with a mean absolute percentage error of 10%, surpassing those simulated using the ray-tracing method with an error margin of 30%, while also reducing computational time. This exceptional fidelity and efficiency empower our model to serve as a cost-effective alternative to synchrotron experiments. Our machine learning-aided simulation method is affordable and readily deployable for a large variety of materials, opening new doors to eliminate or reduce defects for a wide range of laser materials processing techniques.

## 1. Introduction

For over half a century, laser materials processing has been broadly used in our society, including aerospace, automotive, and many other high-tech industries [1, 2]. Defects such as pores formed during laser-material interaction, however, pose a serious threat to the mechanical durability, reliability, and security of these components. For example, the fatigue resistance of a component is significantly decreased due to these defects [3, 4]. Keyhole – a deep and narrow cavity caused by the recoil pressure generated by rapid evaporation – plays a pivotal role in generating defects during the laser materials processing [5]. The fluctuation and collapse of keyholes typically form bubbles and eventually pores in the final products [6, 7]. Yet, the characterization and prediction of keyhole dynamics during laser-material interaction remain a technical challenge because it is a highly localized and ultra-fast process. Recent advancements in high-speed synchrotron X-ray imaging experiments [8, 9] provided insights into keyhole instability under various processing parameters of laser powers ($P$) and scan speeds ($v$) [10, 11]. However, their widespread adoption has been largely impeded by sophisticated instruments and limited access to synchrotron facilities. Hence, there is a compelling need for a low-cost and readily deployable solution to quantify keyhole features for a large variety of processing parameters and materials.

Numerical simulation provides a cost-effective and efficient opportunity to reveal the complex physical mechanisms during laser-metal interaction, including recoil pressure, Marangoni convection, material spattering, and porosity generation [12-16]. However, previous simulations often fail to accurately predict keyhole dimensions and are limited to qualitative analysis [17, 18], which is mainly due to the lack of an accurate heat source model to simulate the input energy of the laser. There are two commonly used heat source models: (i) the surface heat source model with a fixed laser absorptance and (ii) the surface heat model incorporating a multi-reflection ray-tracing method. Laser absorptance, which quantifies the portion of applied laser energy absorbed by the material, is an essential input parameter in simulation models [19-22]. In reported simulation studies, fixed laser absorptance values are selected based on empirical knowledge and used across a large range of *P-v* space, for example, 0.3 for Ti-6Al-4V (Ti64) and 0.7 for pure Al, due to the absence of real-time laser absorptance data [23, 24]. However, it is not rational to use the same empirical laser absorptance for a large variety of processing parameters because keyhole morphologies, especially keyhole depth, are distinct under different parameters, thereby altering real-time laser absorptance [25]. Although the multi-reflection ray-tracing method is theoretically reliable to estimate the dynamic laser absorptance [26, 27], it relies on assumptions about the minimum remaining energy and the maximum number of reflections, causing disparities with experimental values [28]. Additionally, the accuracy of the ray-tracing method depends heavily on the resolution and quality of the computational mesh. As a result, tracking multiple reflections and absorptions in such a high-resolution configuration significantly increases computational time, making it inefficient when investigating keyhole behaviors for a wide range of processing parameters [29, 30].

Recently, machine learning has shown exceptional capability to handle multi-dimensional data and discover implicit relationships within a dataset [31-33]. Therefore, new machine-learning methods have been widely used in monitoring and measuring keyhole features during laser materials processing [34-36]. Nevertheless, the establishment of an accurate prediction model for dynamic keyhole features is still hindered by the absence of a comprehensive dataset such as laser absorptance values across various processing parameters. In this study, we develop a novel framework that integrates a computational fluid dynamics (CFD) model and a machine learning-based laser absorptance model to visualize real-time keyhole morphologies, especially keyhole depth, across a wide range of processing parameters. Compared with previous models reliant solely on theoretical assumptions, we start incorporating experimental laser absorptance data into our simulation model under certain processing parameters for model optimization, as validated by the agreement between simulated and X-ray-observed keyhole morphologies. Moreover, by comparing ray-tracing simulation results with experimental data, it reveals that the accuracy of the ray-tracing method varies significantly with different processing parameters. To address this instability, we derive laser absorptance values using the validated CFD model and readily available experimental X-ray images, generating a dataset to establish a machine learning-based laser absorptance prediction model. Incorporating the predicted laser absorptance into the validated CFD model enables accurate and efficient keyhole depth prediction in a broad $P$-$v$-$r_0$ space ($r_0$ is the laser spot radius on the sample surface).

## 2. Materials and methods

### *2.1 Experimental Investigations on Laser Absorptance and X-ray Image Capture*

A combined integrating sphere and high-speed synchrotron X-ray system was developed to measure laser energy absorptance and capture the X-ray videos at the 32-ID-B beamline of Advanced Photon Source at Argonne National Laboratory. The absorbed laser energy is calculated using an energy balance computation between the measured input and scattered laser light (zero light transmission). To ensure the collection of intense backscattered light from the initially specular surface, the laser was incident at an angle of 7° relative to the sample surface normal. A fiber-coupled photodiode positioned on the sphere surface was used to measure the backscattered light. The resulting photodiode voltage was captured by a high-speed oscilloscope, providing a voltage uncertainty of 1 % and a time resolution of 40 ns. To establish an absolute measurement of the scattered light power, a calibration procedure was conducted using a well-characterized scattering surface instead of the experimental target, enabling the conversion of the photodiode signal into an accurate measurement of the scattered light power. More detailed information about the laser absorptance measurements can be found in Ref. [25]. Experiments were conducted on two $P$ & $v$ combinations for Ti64 and three for Al6061 under a constant surface spot radius of 60 μm on bare metal plates to acquire the real-time laser absorptance data from the first to last moments of laser exposure.

The laser system utilized in the experiments comprises a ytterbium fiber laser and a galvo laser scanner system. The fiber laser has a Gaussian profile with a 1070 nm wavelength and a maximum power output of 540 W. The laser's maximum scanning speed across the sample is 2 m/s. To maintain controlled conditions, the samples are enclosed within a stainless-steel chamber with an argon (Ar) environment at atmospheric pressure (1 atm). During the experiment, the laser interacts with the specimen, while high-energy X-rays penetrate through its thickness. Concurrently, a high-speed camera captures images at 50,000 frames per second. For the X-ray images captured alongside the laser absorptance measurements for both Ti64 and Al6061, the laser spot diameter on the sample surface is 122.5 μm, and the laser spot diameter on the focal plane is 49.5 μm. For the X-ray images collected from the literature for Ti64, the laser spot diameter on the sample surface is 95 μm, and the laser spot diameter on the focal plane is 56 μm. The laser spot diameters on the sample surface for Al6061 are 82.5, 95, and 122.5 μm, respectively. To simplify calculations, we approximated the laser spot radii on the sample surface for simulation models as 60, 50, and 40 μm, respectively. All processing parameters used in this study correspond to the keyhole mode of laser-material interaction.

### *2.2 Data processing and quantification*

To enhance the quantification of keyhole morphologies from the raw X-ray images, we employed a segmentation process to isolate the keyhole area and automatically measure its dimensions, following the procedures outlined in [37]. For the simulation results, we evaluated the keyhole depth and melt pool depth by referencing the isotherms corresponding to the saturation temperature and solidus temperature of the material, respectively, as suggested by Gan *et al*. [38]. To mitigate the influence of volatile fluctuations in keyhole depth calculations, we employed a statistical approach, calculating mean values while excluding the top and bottom 30% of the data points. We illustrated these processes with an example showcasing the raw X-ray images, segmented mask, and simulation results, including keyhole and melt pool contour lines, as shown in Fig. 1.

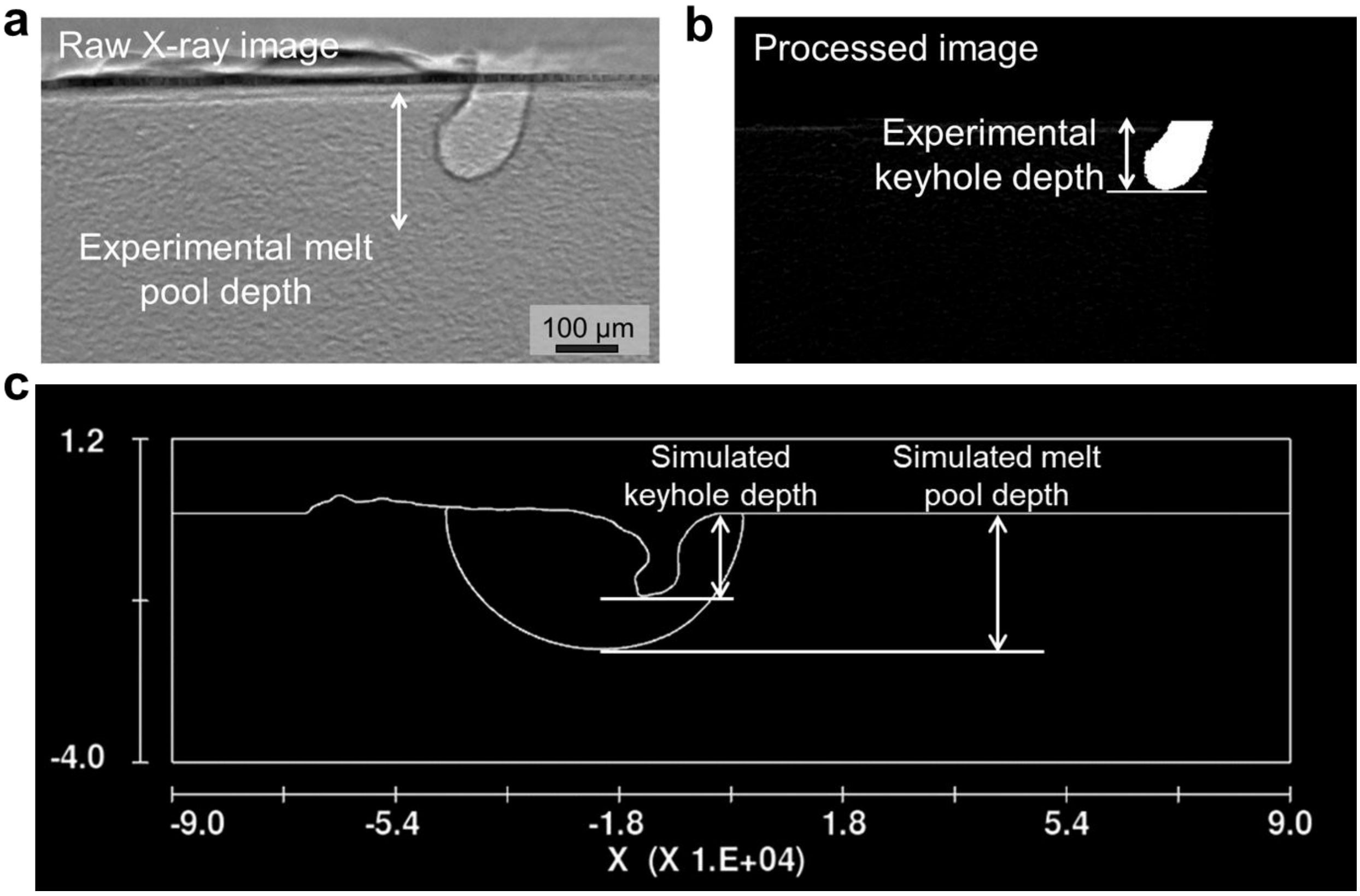

**Fig. 1.** Representative **a** raw X-ray image, **b** segmented mask, and **c** simulated keyhole morphology at the longitudinal cross section. The experimental X-ray images are segmented to isolate the keyhole area and automatically measure the experimental keyhole depth by the method proposed in the Ref. [37]. The simulated keyhole depth and melt pool depth are evaluated by referencing the isotherms corresponding to the saturation temperature and solidus temperature of the material, respectively, as suggested by Ref. [38].

*2.3 Multiphysics thermal-fluid flow model*

In this study, we adopted a multi-physics thermal-fluid flow model, using computational fluid dynamics (CFD) and volume of fraction (VOF) approaches, implemented through Flow 3d v11.2 (Please note that certain commercial products or company names are identified here to describe our study adequately. Such identification is not intended to imply recommendation or endorsement by the National Institute of Standards and Technology, nor is it intended to imply that the products or names identified are necessarily the best available for the purpose). The numerical simulation is based on a set of model assumptions: (i) the liquid in the melt pool is incompressible and Newtonian; (ii) the shielding gas is ignored, and the area other than the fluid is treated as void with uniform temperature and pressure; (iii) phase change is considered while the resulting compositional change is omitted; and (iv) the vapor is not modeled but the effect is considered through recoil pressure [24].

The mass conservation equation, Navier-Stokes equation, and energy conservation equation are given as follows:

$$\nabla \cdot (\vec{v}) = 0 \tag{1}$$

$$\frac{\partial \vec{v}}{\partial t} + (\vec{v} \cdot \nabla) = -\frac{1}{\rho} \nabla Pressure + \mu \nabla^2 \vec{v} - F_d \vec{v} + \vec{G} \tag{2}$$

$$\frac{\partial h}{\partial t} + (\vec{v} \cdot \nabla) h = q + \frac{1}{\rho} \nabla \cdot (k \nabla \mathrm{T}) \tag{3}$$

where $\vec{v}$ ($\mathrm{m \cdot s^{-1}}$) is the velocity vector, $q$ the laser heat source, $t$ (s) the time, $Pressure$ (Pa) pressure, $\rho$ ($\mathrm{kg \cdot m^{-3}}$) density, $\mu$ ($\mathrm{m^2 \cdot s^{-1}}$) viscosity, $h$ ($\mathrm{J \cdot kg^{-1}}$) the enthalpy, and $k$ ($\mathrm{W \cdot m^{-1} \cdot K^{-1}}$) the thermal conductivity. $F_d$ ($\mathrm{s^{-1}}$) is the drag force coefficient and $\vec{G}$ ($\mathrm{m \cdot s^{-2}}$) the body acceleration due to body force.

The primary physics models utilized in simulation encompass laser models and surface forces. For the laser model, a Gaussian heat source is employed to describe the laser energy absorbed by the upper surface and keyhole, as expressed in Equation (4):

$$q = \frac{3P \cdot LA}{\pi \cdot {r_0}^2} e^{\left(\frac{-3(x^2+y^2)}{{r_0}^2}\right)} \tag{4}$$

where $q$ (J· m$^{-2}$·s$^{-1}$) is the laser heat flux absorbed at the free surface at the point ($x$, $y$) and LA is the laser absorptance of the material. A well-established approach from previous publications for determining LA is the ray-tracing algorithm. In this method, the laser beam is dispersed into discrete meshes, each treated as a sub-beam with uniformly distributed energy. The LA of the incoming ray ($I$) energy by the free surface is calculated by the Fresnel absorption [27]:

$$LA = 1 - \frac{1}{2}\left(\frac{1+(1-\varepsilon\cos\varphi)^2}{1+(1+\varepsilon\cos\varphi)^2} + \frac{\varepsilon^2-2\varepsilon\cos\varphi+2\cos\varphi^2}{\varepsilon^2+2\varepsilon\cos\varphi+2\cos\varphi^2}\right) \quad (5)$$

Here, $\varepsilon$ is the material-dependent coefficient and $\varphi$ $(rad)$ is the angle between the laser ray and surface normal unit vector ($N$). The initial point of the reflected ray ($R$) is the sub-cell center, and the vector is calculated by the following equation:

$$R = I - 2\,(I \cdot N)\,N \quad (6)$$

The multiple reflection process will terminate when one of the following conditions is met: (i) the sub-beam exits the keyhole, (ii) the incident angle exceeds π/2, and (iii) the number of reflections of a laser ray exceeds 10, or the remaining energy is less than 0.5% of the original ray energy. Because this study focused on the keyhole dynamics in bare plates, the ray-tracing model here is not designed for scenarios involving the presence of a powder bed.

The heat source is regarded as part of the surface heat flux boundary condition, and the main energy transfer modes in the upper free surface include convection, radiation, and evaporation, which can be expressed using the following equations:

$$k\frac{\partial T}{\partial \vec{n}} = q - q_{conv} - q_{rad} - q_{evap} \quad (7)$$

$$q_{conv} = h_c\,(T - T_{ref}) \quad (8)$$

$$q_{rad} = \sigma\epsilon\,(T^4 - {T_{ref}}^4) \quad (9)$$

$$q_{evap} = \varphi L_v P_{atm}\sqrt{\frac{1}{2\pi RT}}\,exp\left[\frac{L_v(T-T_b)}{TRT_b}\right] \quad (10)$$

Where $\vec{n}$ is the surface normal vector and $h_c$ (W· m$^{-2}$·K$^{-1}$) is the heat transfer coefficient. σ (W· m$^{-2}$·K$^{-4}$) is the Stefan-Boltzmann constant and $\epsilon$ is the radiation emissivity. For other surfaces, only convection and radiation are considered.

Two significant forces act upon the surface of the liquid metal, causing deformation of the free surface. As the heat source is applied, the temperature of the substrate increases, initiating the melting process. Upon reaching the melting point, surface tension predominantly governs the flow behavior. The surface tension coefficient is estimated as a linear function of temperature to account for the Marangoni effect, expressed in the following equation:

$$\sigma\,(T) = \sigma_0 + \sigma_s^T\,(T - T_l) \tag{11}$$

where $\sigma_0$ (N · m$^{-1}$) and $\sigma_s^T$ are surface tension coefficient at the reference temperature $T_l$ (liquidus temperature) and its temperature sensitivity, respectively. When laser energy irradiates the material, it leads to violent evaporation and the generation of a significant amount of vapor, resulting in recoil pressure. The recoil pressure model employed in this study is described by Equation (12):

$$P_r = \frac{1+\beta_R}{2} * P_{atm} * e^{\left(\frac{\Delta H}{R_{gas}} * \left(\frac{1}{T_v} - \frac{1}{T}\right)\right)} \tag{12}$$

where $P_r$ (Pa) is the recoil pressure, $\beta_R$ is the ratio of recondensation particles to the evaporation ones, $P_{atm}$ (Pa) is the ambient pressure, $\Delta H$ (J·mol$^{-1}$) is the specific enthalpy of metal vapor, $R_{gas}$ (J·kg$^{-1}$·K$^{-1}$) is the universal gas constant, $T_v$ (K) is the boiling temperature, and $T$ (K) is the surface temperature. Both surface tension, recoil pressure, and the Marangoni effect are treated as boundary conditions.

The evolution of gas-liquid free surface is tracked by the VOF method:

$$\frac{\partial F}{\partial t} + \nabla \cdot (F\vec{v}) = 0 \tag{13}$$

where $F$ is the volume fraction. The computational domain for the bare substrate in this study spans dimensions of 1800 μm × W μm × ($H$ + 150) μm, where W is 300 μm for Ti64 and 800 μm for Al6061, as shown in Fig. 2. The depth of this computational domain (*H*) is determined based on various input energy density values, ranging from 300 μm to 800 μm. To maintain both simulation accuracy and computational efficiency, the mesh size is set to 6 μm after the mesh-sensitivity analysis. For a comprehensive overview of the thermal properties, which encompass density, thermal conductivity, viscosity, and surface tension, as well as other material properties specific to Ti64 and Al6061 (Fig. 3 and Table 1).

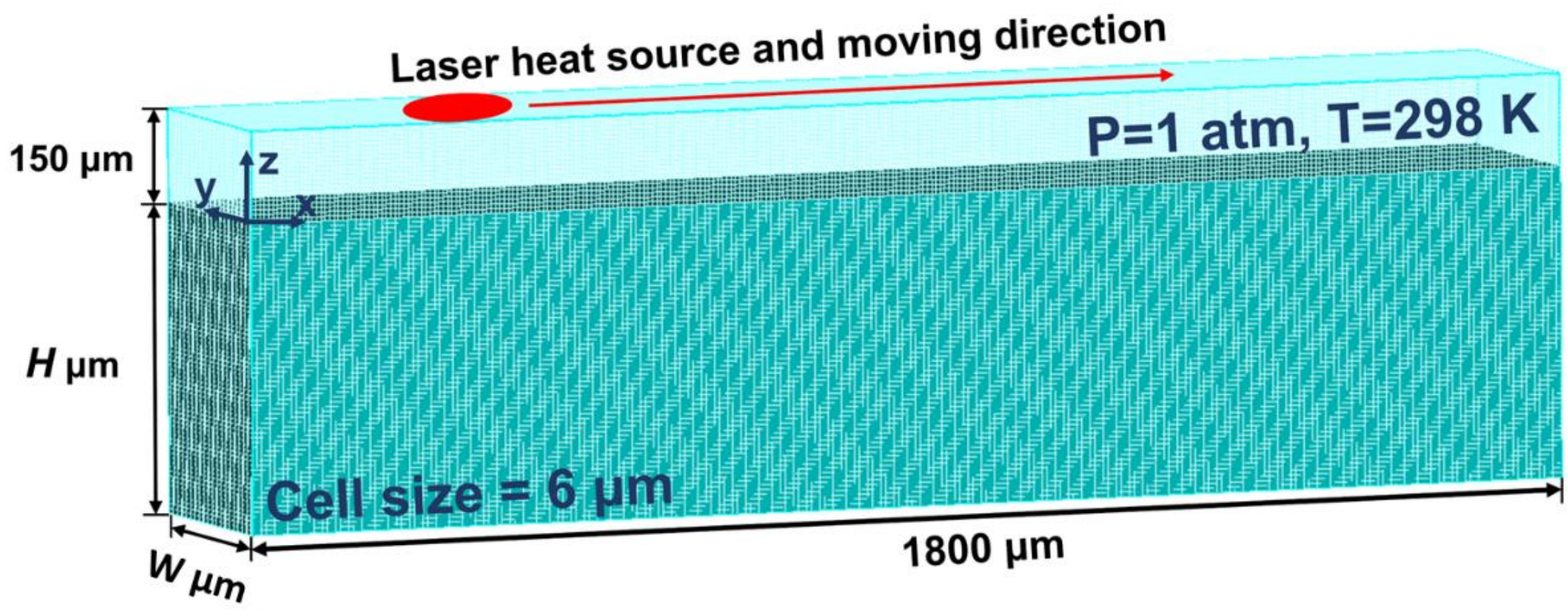

**Fig. 2.** The computational domain for the simulation models, spanning dimensions of 1800 μm × W μm × (*H* + 150) μm, where W is 300 μm for Ti64 and 800 μm for Al6061. The depth of this computational domain (*H*) is determined based on various input energy density values, ranging from 300 μm to 800 μm. After the mesh-sensitivity analysis, the mesh size is set to 6 μm to maintain both simulation accuracy and computational efficiency.

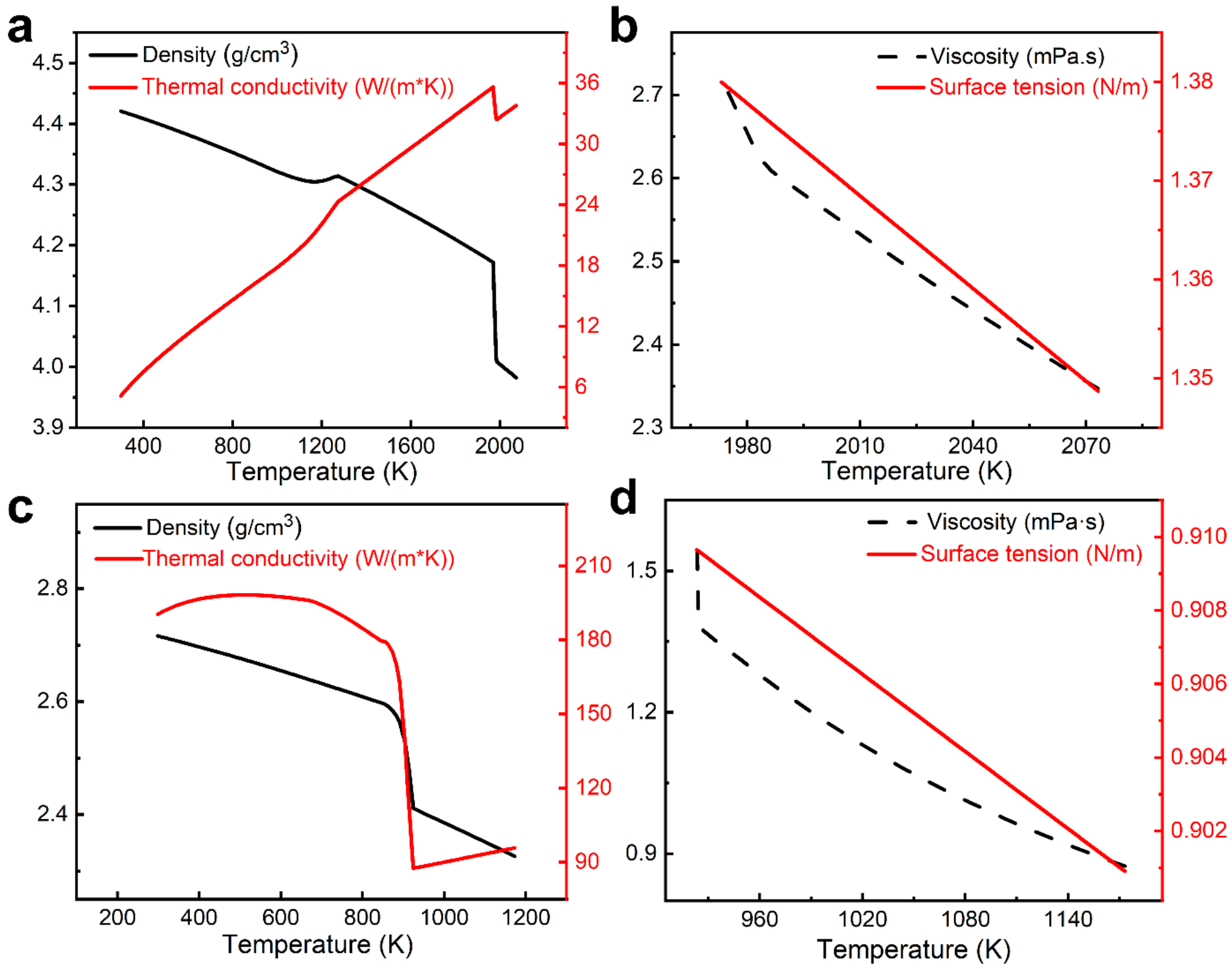

**Fig. 3.** The thermal properties encompassing density, thermal conductivity, viscosity, and surface tension for **a, b** Ti64 and **c, d** Al6061.

**Table 1.** Thermal and mechanical parameters for the simulations of Ti64 and Al6061 [24, 35, 39].

| Property | Ti6Al4V | Al6061 |
|---|---|---|
| Solidus temperature [K] | 1878 | 873.15 |
| Liquidus temperature [K] | 1928 | 915.15 |
| Boiling temperature [K] | 3315 | 2750 |
| Gas constant ($J \cdot kg^{-1} \cdot K^{-1}$) | 173.2 | 308 |
| Latent heat of melting [$J \cdot kg^{-1}$] | $2.86 \times 10^5$ | $3.97 \times 10^5$ |
| Latent heat of evaporation [$J \cdot kg^{-1}$] | $9.7 \times 10^6$ | $1.077 \times 10^7$ |
| Saturated vapor pressure [Pa] | $1.013\times10^{5}$ | $1.013\times10^{5}$ |
| Stefan-Boltzman constant [$W\cdot m^{-2}\cdot K^{-1}$] | $5.6704\times10^{-8}$ | $5.6704\times10^{-8}$ |
| Recondensation coefficient $\beta_R$ | 0.08 | 0.5795 |
| Darcy drag force coefficient | $5.57\times10^{6}$ | $3\times10^{6}$ |
| Surface tension ($N\cdot m^{-1}$) at fusion temperature | 1.38 | 0.91 |
| Thermocapillary coefficient ($N\cdot m^{-1}\cdot K^{-1}$) | $-3.13\times10^{-4}$ | $-3.5\times10^{-4}$ |
| Surface radiation coefficient $\varepsilon$ | 0.4 | 0.15 |

### *2.4 Machine learning models*

We selected and employed six commonly used machine learning-based regression models to predict laser absorptance under varying processing parameters. We implemented a 5-fold nested cross-validation technique to train these models. For this study, the hyperparameters of all selected models were fine-tuned through a Bayesian optimization algorithm with an acquisition function of expected improvement via a commercially available software regression learner toolbox [40]. The first regression model we selected was linear regression (LR) [41] due to its simplicity, adaptability, and computational efficiency. To address noise and enhance robustness, we used the support vector regression (SVR) model [42] and further optimized its kernel function and corresponding scale values. Moreover, the decision tree (DT) [43] model was chosen to capture non-linear relationships, and we employed random forest (RF) as an ensemble method to mitigate overfitting and instability [44]. We hyper-tuned the minimum leaf size and number of learners for the RF model. Artificial neural network (ANN) [45] was included for its ability to analyze intricate nonlinear relationships, optimized by tuning the number of layers and layer size. Lastly, the Gaussian process regression (GPR) model [46] was included due to its usability and flexibility in implementation, with optimization of kernel functions and scale values.

## 3. Results and discussion

We predict keyhole dynamics using a multi-physics CFD model, enhanced by different approaches for determining laser absorptance. In section 3.1, we validate that the CFD model using experimentally measured laser absorptance values can reliably reproduce experimental keyhole morphologies and depth. In Section 3.2, we assess the performance of the widely used ray-tracing method for predicting laser absorptance as input to the CFD model, highlighting its limitations through comparisons with synchrotron-based keyhole data from the literature. To improve prediction fidelity, Section 3.3 introduces a physics-based inverse approach that iteratively refines absorptance inputs based on Gan's analytical theory [38], though its practical implementation faces accuracy limitations and divergence issues. Finally, Section 3.4 presents the development of a machine learning model that directly predicts absorptance from processing parameters, achieving superior accuracy and efficiency compared to traditional methods.

### *3.1 Multi-physics simulations based on real-time laser absorptance measurements*

The experimentally measured laser absorptance initially increases and gradually stabilizes, for example, after 0.4 ms for Ti64 and 0.6 ms for Al6061, respectively (Fig. 4a and 4b). The initial high absorption is a start-of-line feature due to keyhole initiation under sufficiently high laser irradiation [25, 47]. Within the keyhole cavity, the laser undergoes multiple reflections, leading to an increase in the laser absorptance compared to that on a flat surface [20]. Meanwhile, a pronounced Marangoni effect [48] transports hotter molten metal toward colder regions, resulting in a reduction of the keyhole depth and multiple reflections, which lowers the laser absorptance. Under these conditions, the combined effects of metal vaporization and fluid dynamics maintain the equilibrium of keyhole morphologies, as elucidated in [25]. In this work, we focus on the period of stable melting, as indicated by the regions between dashed lines (Fig. 4a and 4b). Table 2 shows the calculated mean values of the real-time laser absorptance during stable melting for all five selected processing parameters. There is a significant divergence between experimental results and predictions using the ray tracing approach [27], suggesting the limitations of using this commonly used mode to obtain laser absorptance.

**Table 2.** The mean values of the real-time laser absorptance (LA) during stable melting for Ti64 and Al6061 bare plates under five processing parameters.

| Material | Laser power (W) | Scan speed (m/s) | Mean LA measured by the experiment | Mean LA predicted by ray-tracing [27] |
|---|---|---|---|---|
| Ti64 | 200 | 0.7 | 0.50 | 0.61 |
| Ti64 | 254 | 0.7 | 0.64 | 0.72 |
| Al6061 | 557.5 | 0.7 | 0.56 | 0.45 |
| Al6061 | 500 | 0.7 | 0.49 | 0.4 |
| Al6061 | 472.6 | 0.7 | 0.45 | 0.32 |

To predict the keyhole depths, we adopted a multi-physics thermal-fluid flow model, using CFD with a volume of fluid (VOF) approach. We compared the time-resolved keyhole depth obtained from simulations and experimental results derived from X-ray images. Our simulation results of keyhole morphologies are comparable to those observed in experimental X-ray images (Fig. 4e-h): the simulated keyhole depths (Ti64: 54 ± 4 μm; Al6061: 143 ± 21 μm) versus experimental keyhole depth (Ti64: 55 ± 5 μm; Al6061: 148 ± 29 μm). Compared to Ti64, Al typically necessitates a higher energy density input for processing, primarily because of its larger laser reflectivity and the presence of supercritical oxide [49]. This increased energy density input facilitates the formation of a higher aspect ratio keyhole, boosting the fluctuation frequency of the keyhole [10]. Therefore, the laser absorptance during the stable period of Al6061 exhibits more severe fluctuations than that of Ti64. Moreover, for all five data points (Table 2), our simulation results match the experimental results well (Fig. 5), demonstrating the consistently high performance of our model (Videos 1-5).

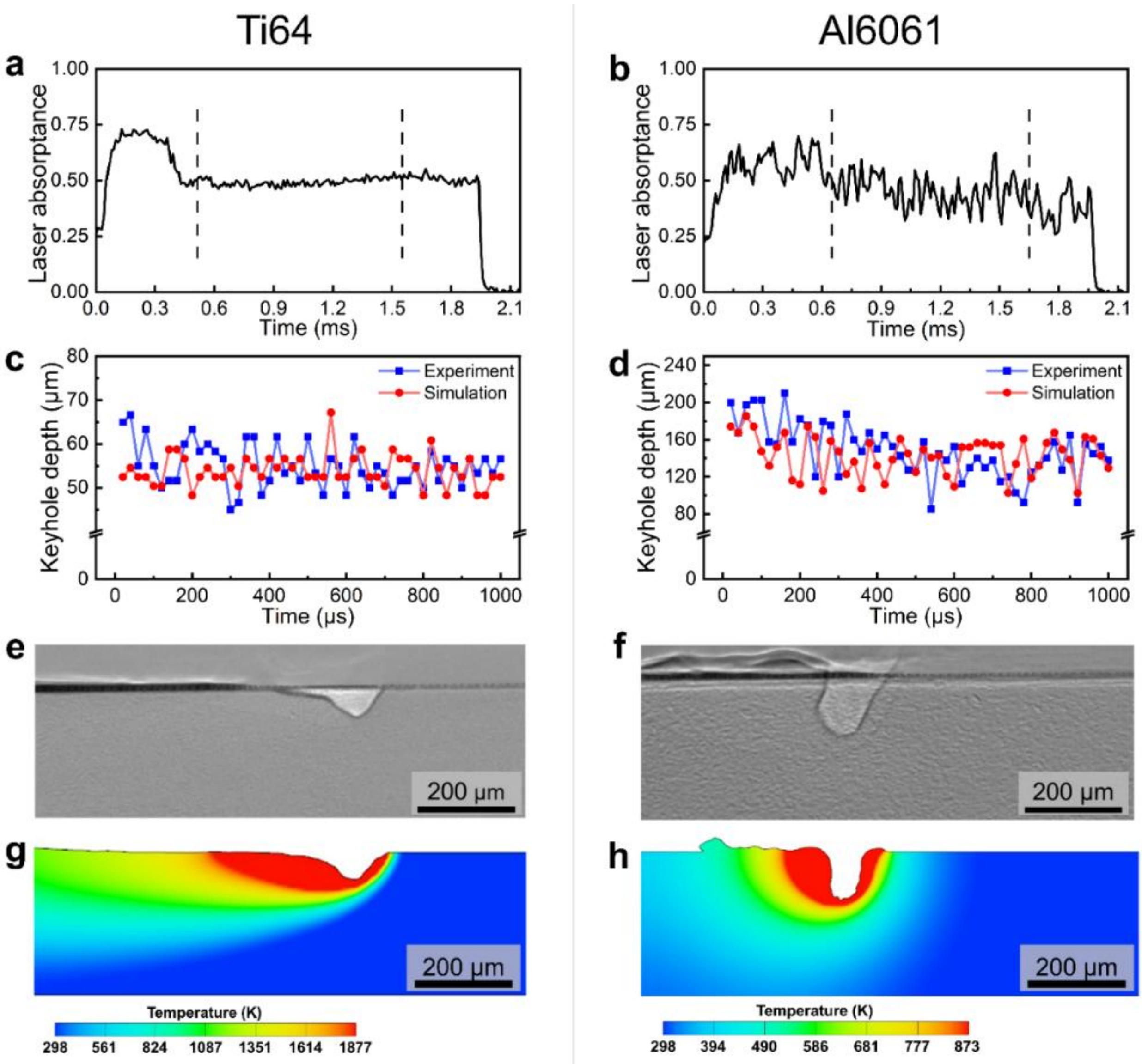

**Fig. 4.** Multi-physics simulations for Ti64 under a laser power of 200 W and Al6061 under a laser power of 500 W, using the experimental laser absorptance. **a, b** Experimentally measured real-time laser absorptance. The steady-state conditions used for the analysis of keyhole depths are indicated by dashed lines. **c, d** Comparison of keyhole depths generated by our simulation and measured by experiments. This comparison is conducted with a time interval of 1 ms after keyhole fluctuations had reached a plateau. **e, f** Two selected X-ray images showing the keyhole morphologies for Ti64 and Al6061, respectively. **g, h** Simulation results showing temperature contours to compare to **e** and **f**, respectively. The simulated keyhole morphologies match the experimental observations well.

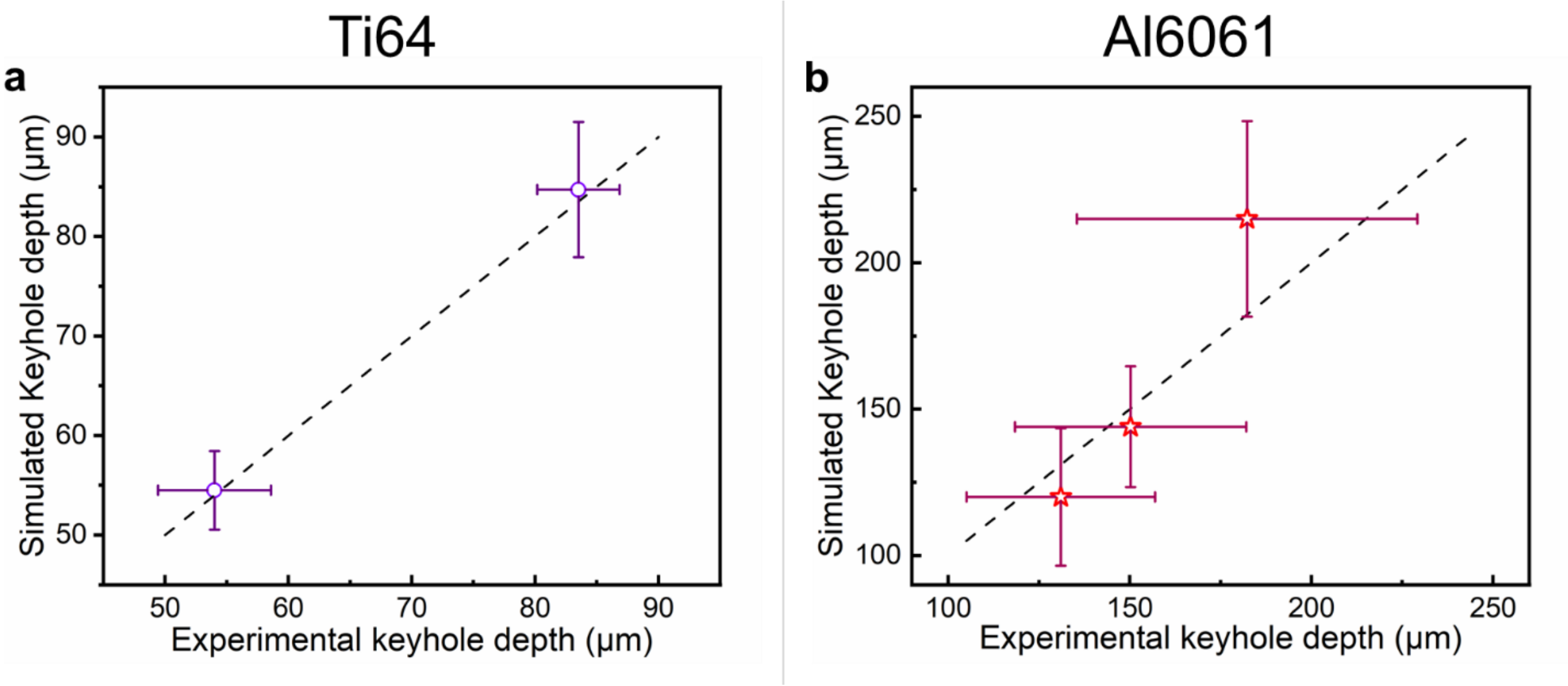

**Fig. 5.** A comparison between the experimental keyhole depth derived from X-ray images and the simulated keyhole depth using experimental laser absorptance for **a** Ti64 and **b** Al6061.

*3.2 Limitations of the ray-tracing method and laser absorptance dataset generation*

In this study, we collected the synchrotron X-ray images of Ti64 and Al6061 processed on bare metal plates under a wide range of laser scanning parameters, as reported in the literature [6, 10] (Fig. 6). We examined the accuracy of the ray-tracing method in terms of predicting the keyhole depth across all the collected processing parameters. Fig. 7 shows the comparisons between experimental and simulated results for Ti64 and Al6061 under various processing parameters. The average error in Al6061 keyhole depth simulation reaches 50%, indicating the poor accuracy of the ray-tracing method. Although the simulation for Ti64 keyhole depth is more accurate than that for Al6061, the average error of Ti64 remains as high as 20%, which is still insufficient. More notably, the accuracy of the ray-tracing method varies significantly under different processing parameters, with discrepancies ranging from 3% to 50%. The unpredictable and inaccurate performance of the ray-tracing method motivates us to establish a more accurate and robust model to predict the laser absorptance values based on the processing parameters. However, the prerequisite of establishing such a model necessitates a dataset with laser absorptance values across various processing parameters.

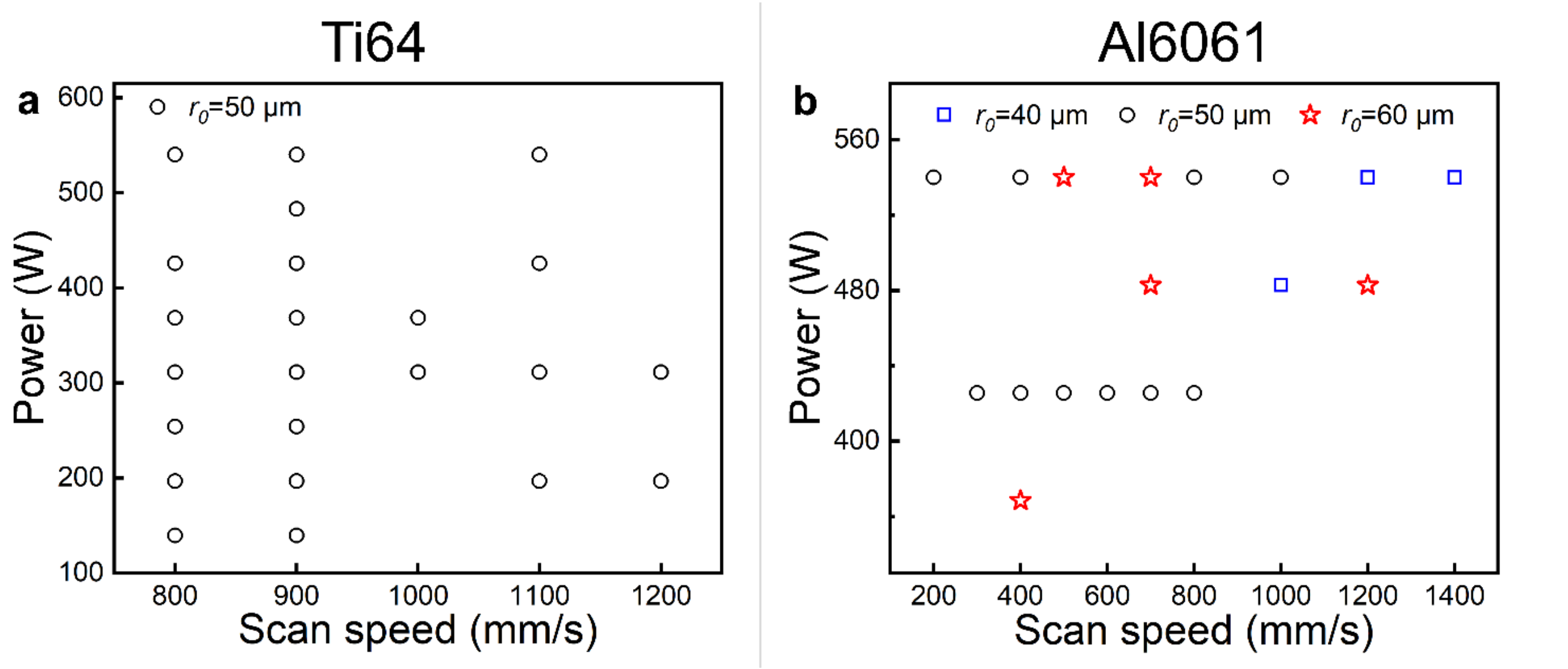

**Fig. 6.** The dataset comprising experimental X-ray images, obtained across a wide range of power ($P$), velocity ($v$), and laser spot radius on the sample surface ($r_0$) combinations. **a,** Data points for Ti64. **b,** Data points for Al6061.

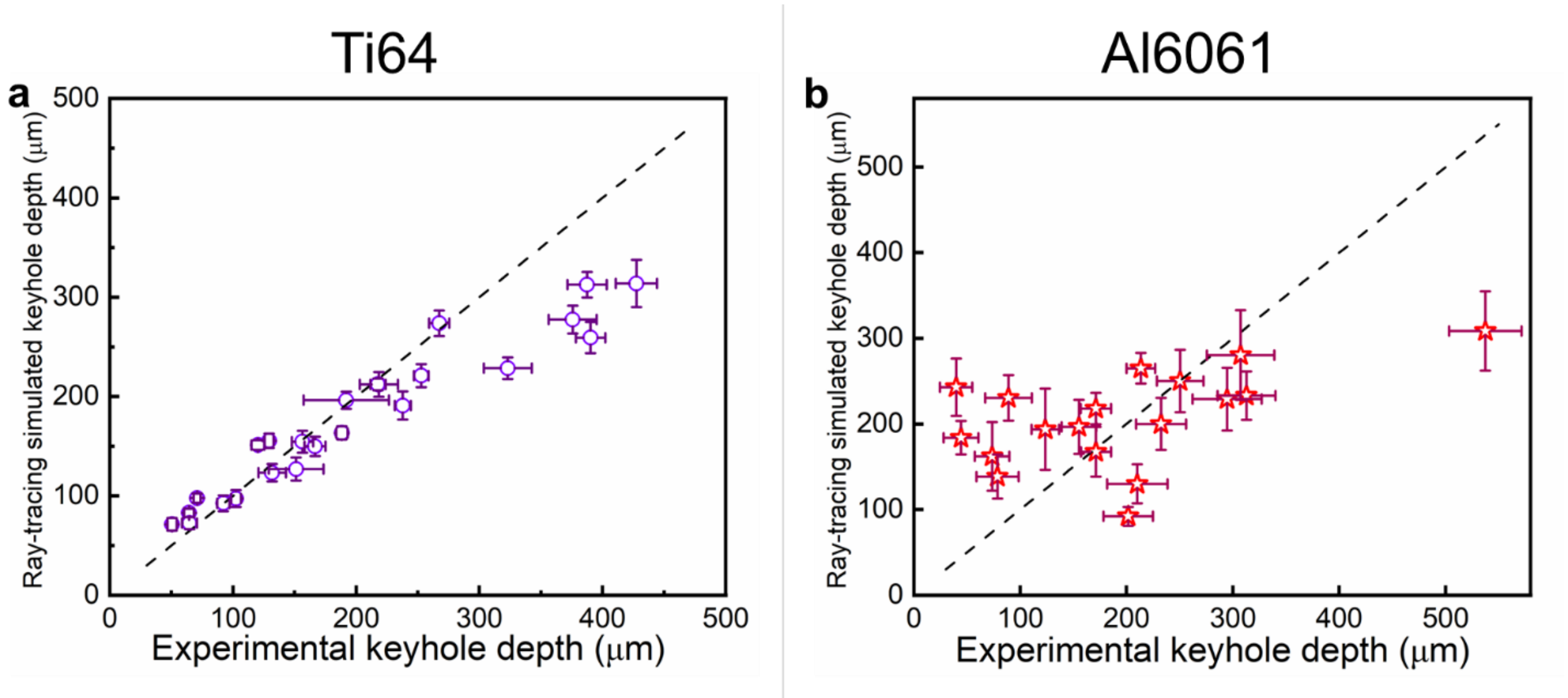

**Fig. 7.** Comparison between keyhole depth from experiments and keyhole depth simulated by the Ray-tracing method for **a** Ti64 and **b** Al6061 under a wide range of processing parameters. The ray-tracing method exhibits unpredictable behavior and frequent inaccuracies in keyhole depth prediction.

Therefore, we adjusted the laser absorptance values in CFD models to match the simulated keyhole depth with the experimentally measured value. This process leads to a compiled dataset of laser absorptance for a set of 46 processing parameters (25 for Ti64 and 21 for Al6061), including both experimental and derived values (Supplementary Data 1). Visual representations of three selected data points under different input energy densities exhibit good agreement in keyhole morphologies between the simulated and experimental results for both Ti64 (Figs. 8a1-a3) and Al6061 (Figs. 8b1-b3). The simulation model not only accurately predicts keyhole depth but also effectively captures other keyhole features that were not used to derive the absorptance values. Moreover, a comparison between the simulated Al6061 melt pool depth and experimental values analyzed from the X-ray images validates the accuracy of the derived laser absorptance (Fig. 9). Our results indicate the feasibility of accurately deriving laser absorptance from X-ray images of the keyhole.

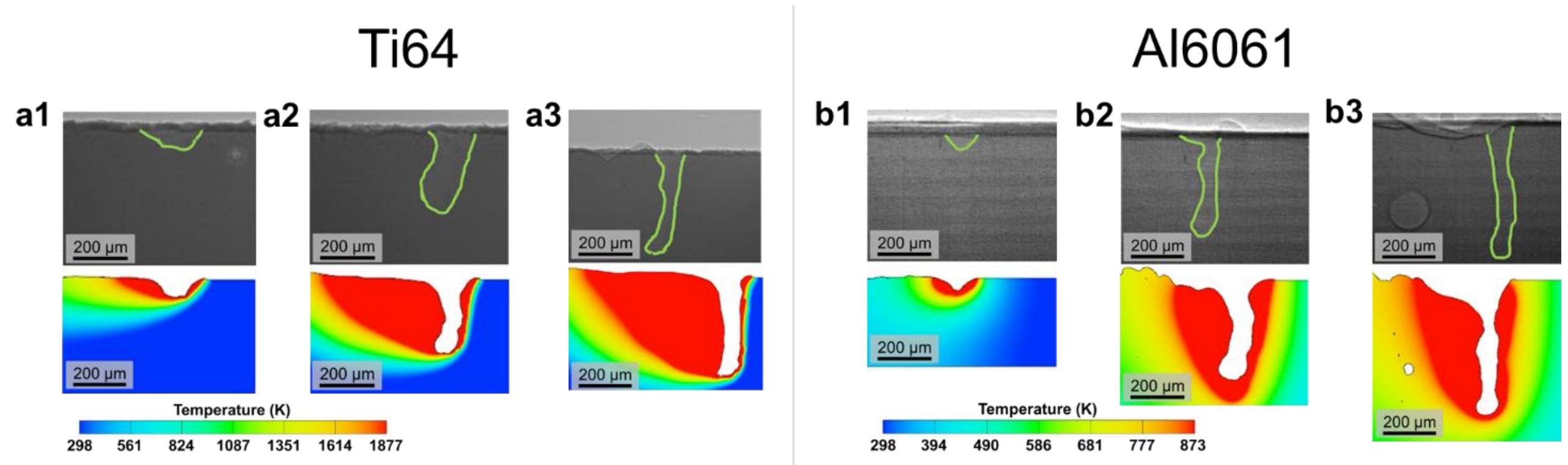

**Fig. 8.** Laser absorptance derivation case studies for Ti64: **a1** *P*: 140 W, *v*: 0.9 m/s, and $r_0$: 50 µm; **a2** *P*: 368 W, *v*: 0.8 m/s, and $r_0$: 50 µm; **a3** *P*: 540 W, *v*: 0.8 m/s, and $r_0$: 50 µm. Laser absorptance derivation case studies for Al6061: **b1** *P*: 540 W, *v*: 0.8 m/s, and $r_0$: 50 µm; **b2** *P*: 425 W, *v*: 0.3 m/s, and $r_0$: 50 µm; **b3** *P*: 540 W, *v*: 0.2 m/s, and $r_0$: 50 µm.

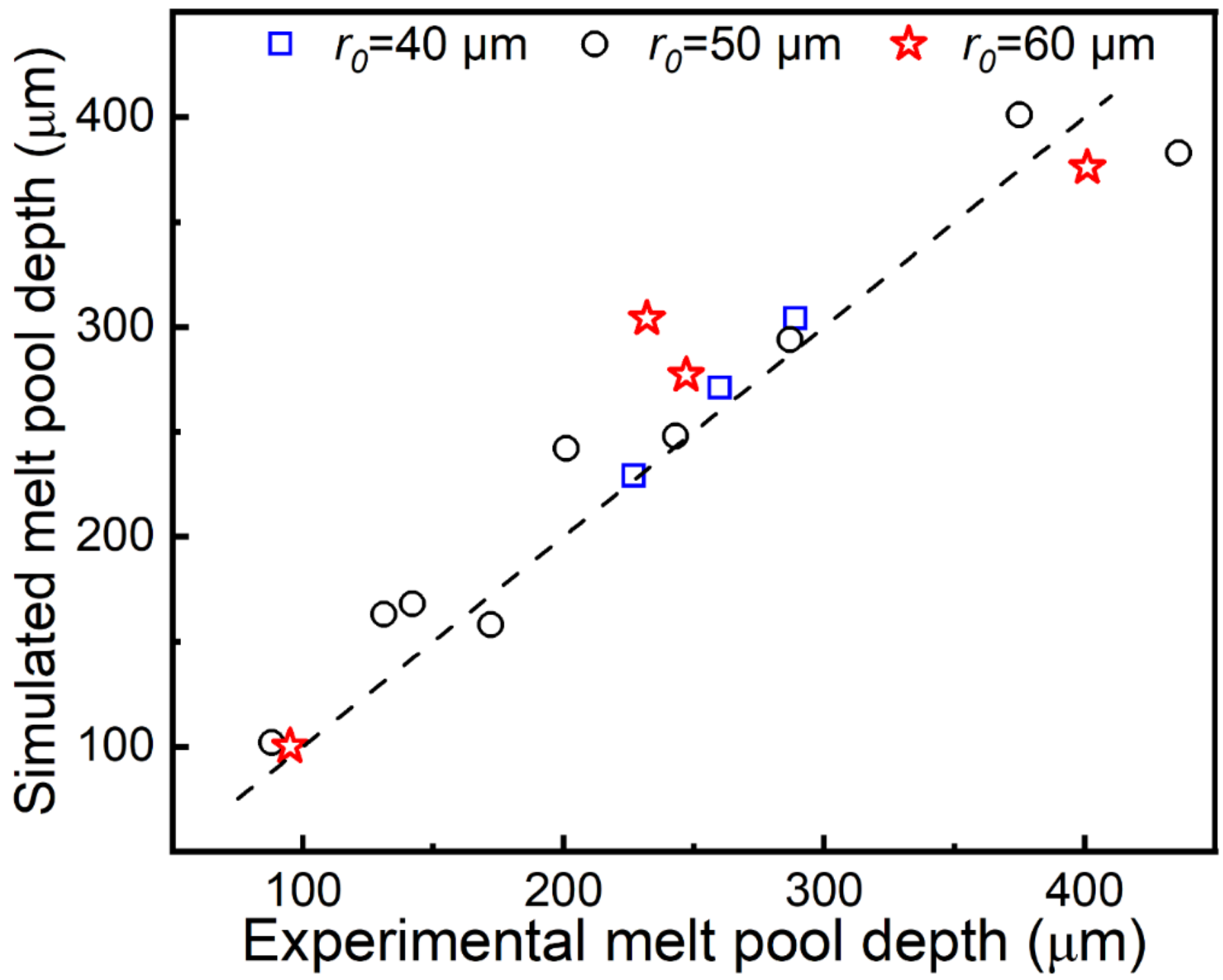

**Fig. 9.** Comparison between experimental melt pool depth and simulated melt pool depth under different surface energy densities for Al6061. The experimental melt pool depths are manually measured from X-ray images for Al6061 due to its distinct contrast.

*3.3 A physics-based approach for keyhole depths prediction*

Based on the generated dataset, we developed two new approaches to predict laser absorptance under new processing parameters for Ti64 and Al6061: a physics-based approach and a machine learning-based approach. The physics-based approach resolves the laser absorptance and keyhole depth using physical models that integrate a forward simulation model (*SIM*) with a backward analytical model (*ANA*). The forward simulation model is used to predict the keyhole depth given the laser absorptance (LA) and the processing parameters ($P$, $v$, and $r_0$), while the backward analytical model is designed to predict the laser absorptance value based on the keyhole depth (KD) and the processing parameters. The iterative solutions of laser absorptance and keyhole depth are calculated using the following equations:

$$Equations: \begin{cases} KD = SIM(P, v, r_{0,} LA) \\ LA = ANA(KD, P, v, r_{0,}) \end{cases} \tag{14}$$

$$Solve: ANA\left(P, v, r_{0,}, SIM\left(LA, P, v, r_{0,}\right)\right) = LA \tag{15}$$

The backward analytical model is approximated using a linear regression function between laser absorptance and a dimensionless variable $X$, drawing on Gan's work [38]:

$$X = \frac{KD\cdot(T_l - T_0)\cdot\pi\cdot\rho\cdot C_p\cdot\sqrt{\alpha\cdot v\cdot r_0}}{P} \tag{16}$$

which are calculated using liquidus temperature $T_l$ (K), substrate temperature $T_0$ (K), density $\rho$ ($g\cdot cm^{-3}$), heat capacity $C_p$ ($J\cdot K^{-1}$), thermal diffusivity $\alpha$ ($m^2\cdot s^{-1}$), and keyhole depth KD (m). This backward analytical model is trained using the generated dataset. In laser materials processing, the $r_0$ also plays a vital role in the melting and evaporation of the material [50]. This motivated the adaptation of the backward analytical models (between laser absorptance and $X$) under different $r_0$ (Fig. 10a and 10b).

Fig. 10c shows that the iterative process converged during the fifth iteration when using the physics-based approach ($P$: 196 W, $v$: 1 m/s, and $r_0$: 50 μm for Ti64) and we observed a 36% discrepancy between the simulated and experimental keyhole depths (95 ± 6 μm vs. 72 ± 3 μm). This discrepancy is mainly due to the moderately linear correlation between laser absorptance and $X$, as indicated by a Pearson coefficient of 0.87 (Fig. 10a). This moderate linear relationship indicates that a significant degree of error persists during the iterations of the laser absorptance prediction. For Al6061, the physics-based approach, when applied to new processing parameters ($P$: 540 W, $v$: 0.6 m/s, and $r_0$: 50 μm), fails to converge (Fig. 10d). Our results suggest that the physics-based approach is theoretically applicable, but the practical challenges posed by the linear approximation and divergence issue hinder its accurate prediction of keyhole depths.

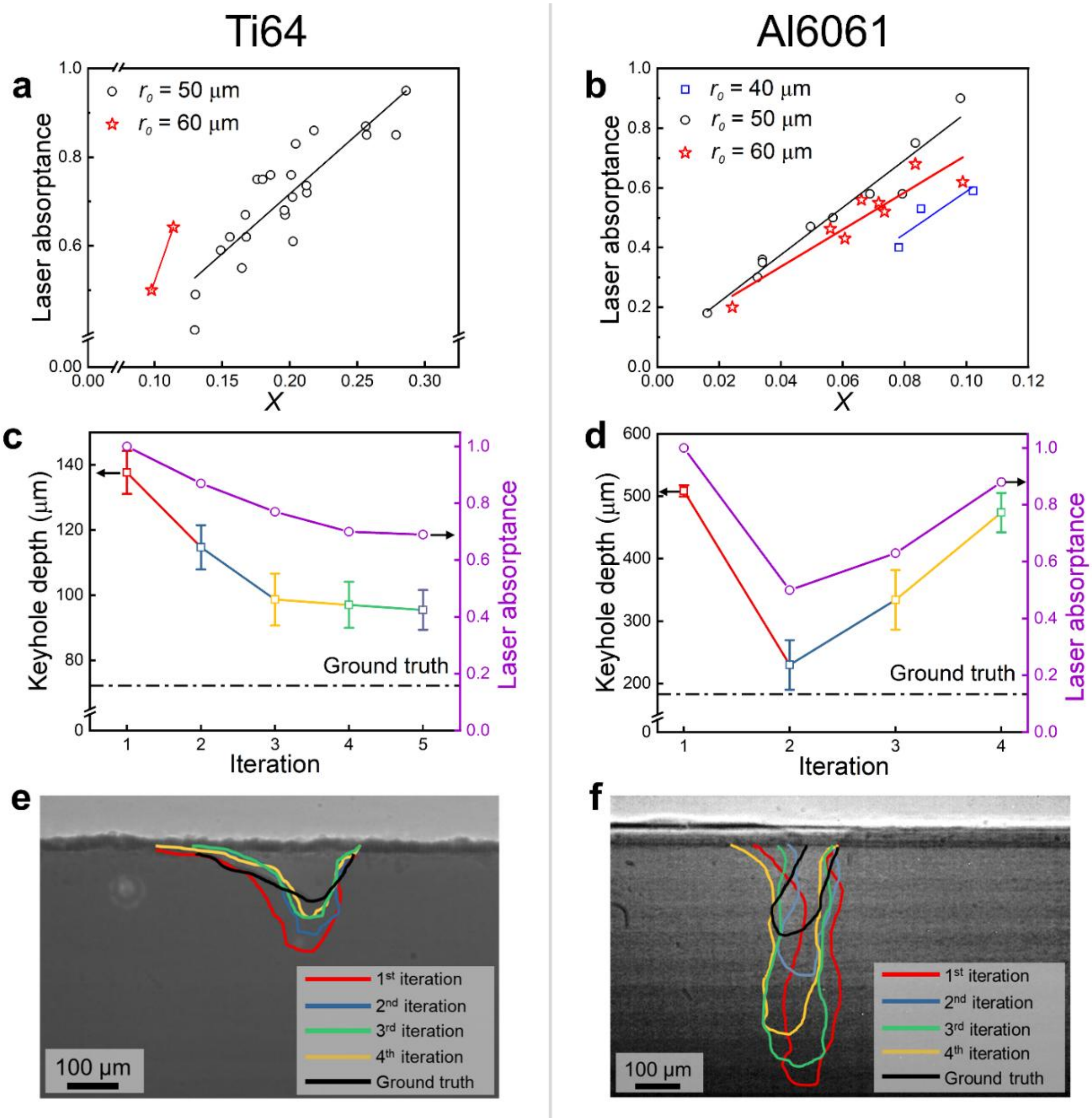

**Fig. 10.** A physics-based approach designed to predict the laser absorptance and keyhole depth. Analytical models for **a** Ti64 and **b** Al6061 are constructed using a linear relationship between the laser absorptance and a dimensionless variable $X$. **c** A case study of Ti64 conducted under selected processing parameters ($P$: 196 W, $v$: 1 m/s, and $r_0$: 50 μm). **d** A case study of Al6061 conducted under selected processing parameters ($P$: 540 W, $v$: 0.6 m/s, and $r_0$: 50 μm). **e, f** Simulated keyhole shape for each iteration overlaid on the raw X-ray images with distinct colors. The experimental keyhole shape is outlined with the black line. The results show that the physics-based approach fails to predict the keyhole morphologies accurately, especially for Al6061.

*3.4 A machine learning-based approach for keyhole depths prediction*

We used the machine learning-based approach to predict laser absorptance based on processing parameters. The predicted laser absorptance value is subsequently incorporated into the established simulation model to predict the keyhole dynamic behavior. To select the appropriate machine learning model, we consider six classic regression models: linear regression (LR), support vector regression (SVR), decision tree (DT), random forest (RF), artificial neural network (ANN), and Gaussian process regression (GPR). The performance of all the regression models is evaluated using the mean absolute percentage error (MAPE) metric and root mean squared error (RMSE) according to the following equations:

$$MAPE = \frac{1}{n}\sum_{t=1}^{n}\left|\frac{Y_{PRE_t}-Y_{EXP_t}}{Y_{EXP_t}}\right| \tag{17}$$

$$RMSE = \sqrt{\sum_{i=1}^{n}\frac{(Y_{PRE_t}-Y_{EXP_t})^2}{n}} \tag{18}$$

where $Y_{PRE_t}$ and $Y_{EXP_t}$ are the predicted and experimental laser absorptance for the $t_{\mathrm{th}}$ sample, respectively.

Table 3 shows the performance of all the selected regression models. Among all the models, GPR demonstrates the strongest overall predictive performance, particularly for Al6061, where it achieves the lowest MAPE and RMSE. For Ti64, while GPR offers the lowest mean MAPE, its RMSE is marginally higher than that of the RF model. Considering its consistently high performance across both materials, GPR was selected as our choice for the predictive model. The GPR models for laser absorptance prediction are accessible via this link: https://github.com/Barry-ZhangUofT/ML-model-for-the-LA/tree/main. Compared to Ti64, the predictive accuracy of the GPR model for Al6061 laser absorptance is lower, which could be attributed to two main factors: (i) The dataset for Al6061 laser absorptance is smaller and spans a broader design space with varying focal spot sizes ($r_0$), whereas for Ti64, all $r_0$ values are fixed at 50 microns. However, this uneven distribution of the Ti64 dataset also limits the generalizability of the prediction model for Ti64 laser absorptivity. (ii) During the laser-material interaction, Al6061 exhibits more severe keyhole fluctuations than Ti64, causing greater variations in laser absorption over time. Because the machine learning model is designed to predict the mean laser absorptance, it does not effectively capture these dynamic fluctuations. As synchrotron data become more widely available in the literature, the accuracy of our model can be systematically improved by incorporating these new data points. In comparison, the current well-established ray-tracing method could only be enhanced through the improvement of its theoretical framework.

**Table 3.** The Mean Absolute Percentage Error (MAPE) and Root Mean Squared Error (RMSE) values for the six regression models in predicting laser absorptance (LA). The GPR model is finally adopted due to its consistently high performance across both materials.

| Models | MAPE for LA (%) | | RMSE for LA (%) | |
|---|---|---|---|---|
| | Ti64 | Al6061 | Ti64 | Al6061 |
| LR | 9.4 ± 4.5 | 31.8 ± 11 | 7.3 ± 1.7 | 14.3 ± 2.1 |
| DT | 9.6 ± 2.7 | 31.7 ± 16 | 9.4 ± 3.6 | 15.1 ± 3.8 |
| RF | 8.6 ± 1.3 | 26.5 ± 13 | 6.2 ± 1.9 | 13.6 ± 4.5 |
| ANN | 10.4 ± 5.6 | 25.2 ± 14 | 8.8 ± 3.1 | 10.6 ± 4.3 |
| SVR | 9.8 ± 3.5 | 29.3 ± 16 | 8.4 ± 1.6 | 16.1 ± 7 |
| GPR | 7.6 ± 4 | 11.5 ± 4.5 | 6.9 ± 0.9 | 5.9 ± 1.4 |

Subsequently, we incorporated both the laser absorptance predicted by the GPR model and laser absorptance from the ray-tracing method into the simulation model to predict the keyhole depths under identical processing parameters as the physics-based approach. The machine learning-based approach achieves the highest accuracy in predicting keyhole depths with the shortest time of implementation, compared with the other approaches (Fig. 11). The ground truth of keyhole depth is analyzed from time-resolved experimental X-ray images. The simulated keyhole depths within a time interval of 1 ms from all three approaches (i.e., ray-tracing, physics-based, and machine learning-based) are illustrated (Figs. 11a and 11b). Furthermore, we show visual representations of keyhole morphology simulated by these three approaches, paired with X-ray images (Fig. 11a1-a4 and 11b1-b4), suggesting that the machine learning-based approach accurately replicates the ground truth of keyhole morphology. Our videos also show the virtualized comparison between experimental X-ray videos and simulated keyhole fluctuations, indicating that our machine learning-based approach best matches the experimental results (Videos 6 and 7).

Another noteworthy point is computational power consumption across different approaches. In the study of Ti64 with a total mesh count of 1.2 million, the machine learning-based approach completes the process in 100 minutes on a desktop with an AMD 7950X3D CPU, while the ray-tracing approach requires 20% more computation time. The physics-based approach requires four rounds iterative simulations to complete, making it at least four times as slow as the other two approaches. In the Al6061 case study, where the total grid count increases to nearly 8 million and keyhole fluctuations become more pronounced, the machine learning-based approach takes 26 hours to complete. The ray-tracing method requires approximately 40 hours—an increase of about 54% in computation time. The physics-based approach fails to converge after four iterative simulations and is thus excluded from consideration. Moreover, as the grid count increases, the computational time gap between the machine learning-based and ray-tracing approaches is expected to widen further. The superior accuracy and efficiency of the machine learning-based approach make it a more suitable candidate for studying keyhole behaviors across various processing parameters and material systems.

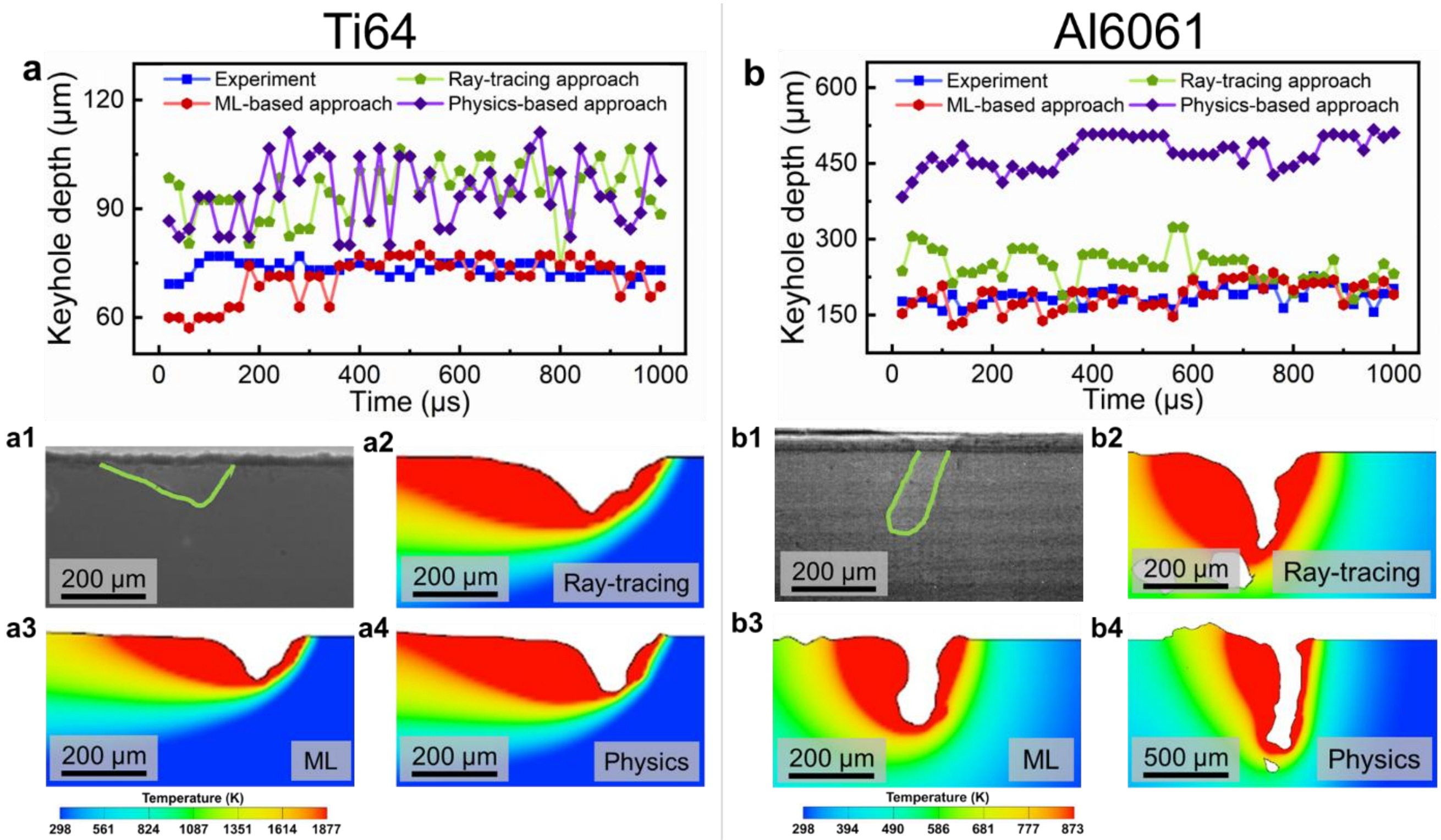

**Fig. 11.** Comparative analysis of three methods for keyhole instability prediction against experimental results for Ti64 ($P$: 196 W, $v$: 1 m/s, and $r_0$: 50 μm) and Al6061 ($P$: 540 W, $v$: 0.6 m/s, and $r_0$: 50 μm). **a, b** Comparison of keyhole depths simulated by three approaches and measured in experiments over a time interval of 1 ms. X-ray images showing keyhole morphologies for **a1** Ti64 and **b1** Al6061. Visualization of keyhole morphologies for **a2-a4** Ti64 and **b2-b4** Al6061, simulated by the method from ray-tracing approach, machine learning-based (ML) approach, and physics-based approach, respectively.

## 4. Conclusions and outlook

Overall, by incorporating the CFD model with the experimental laser absorptance, we accurately predict the real-time keyhole depth for Ti64 and Al6061 on bare metal plates under various $P$, $v$, and $r_0$ processing conditions, maintaining accuracy within a 10% margin. Leveraging the reliability of the CFD model, we generate a dataset comprising laser absorptance from 25 $P$-$v$-$r_0$ combinations for Ti64 and 21 $P$-$v$-$r_0$ combinations for Al6061. Subsequently, a GPR model is selected and trained based on this dataset to predict laser absorptance under new processing parameters. Our method, by using laser absorptance predicted by the GPR model, leads to improved accuracy, efficiency, and robustness in predicting keyhole depth compared to the commonly used ray-tracing approach. Our model's high accuracy eliminates the need for costly and labor-intensive experiments under all possible processing conditions. This study also provides a valuable alternative for researchers without access to synchrotrons, enabling them to improve their model predictions. Furthermore, our model reduces computational cost and time, increasing the efficiency of exploring keyhole behaviors in materials across a broad design space.

To broaden the applicability of the simulation model, there are several potential improvements: (i) Although the simulation model has achieved a decent accuracy, the available laser absorptance data remains limited, necessitating additional experimental laser absorptance values or X-ray data across different $P$-$v$-$r_0$ combinations. (ii) Extending current laser absorptance prediction models to encompass a broader range of materials commonly utilized in laser processing technologies will significantly broaden the application scenarios of the model. (iii) Our current machine learning model for laser absorptance value is limited to laser processing on bare metal plates. The next step is to enhance this model by incorporating additional factors, such as the presence of a powder-bed layer, that influence laser absorptance [51], therefore broadening its applications towards laser-based additive manufacturing. These potential improvements aim not only to enhance the accuracy and versatility of the simulation model but also to contribute to advancing our fundamental understanding and control of laser-based techniques across a wide range of materials and processing parameters.

**Acknowledgments**

J.Z. and Y.Z. acknowledge the financial support from the Centre for Analytics and Artificial Intelligence Engineering (CARTE) Seed Funding program, Data Sciences Institute Catalyst Grant, and NSERC Alliance Grants—Missions ALLRP 570708-2021. This research is part of the University of Toronto's Acceleration Consortium, which receives funding from the Canada First Research Excellence Fund (CFREF). The authors thank V. Malave, N. Tomlin, and J. Lehman from NIST for their helpful comments on the manuscript. J.Z. acknowledges the valuable suggestions for the preparation of figures and videos from Y. Lyu from the University of Toronto. B.S. acknowledges resources of the Advanced Photon Source; a U.S. Department of Energy (DOE) Office of Science user facility operated for the DOE Office of Science by Argonne National Laboratory under Contract No. DE-AC02-06CH11357. The work acknowledges Prof. T. Sun in designing and implementing the apparatus used to obtain the high-speed x-ray images used herein.

**Competing interests:** All authors declare no financial or non-financial competing interests.

**Data and materials availability:** The time-resolved laser absorptance and the X-ray images acquired at the same time under a laser power of 200 W and a scan speed of 0.7 m/s are available on the NIST Public Data Repository (Simonds et al., 2024). Other absorption datasets can be made available upon reasonable request to B.J.S. The X-ray dataset can be made available upon reasonable request to A.D.R. The dataset of segmented keyhole images can be made available upon reasonable request to J.Z. Additional data, including the codes, are available from the corresponding author upon reasonable request.